\pdfoutput=1
\documentclass[aps,prd,reprint,twocolumn,showpacs,superscriptaddress,nofootinbib]{revtex4-1}  
\usepackage{tabularx}
\makeatletter
\def\hlinewd#1{%
\noalign{\ifnum0=`}\fi\hrule \@height #1 %
\futurelet\reserved@a\@xhline}
\makeatother
\usepackage{scrextend}
\usepackage{amsmath}
\usepackage{amssymb}
\usepackage{epsfig}
\usepackage{graphicx}
\usepackage{hyperref}
\usepackage{dcolumn}   
\usepackage{slashed}
\usepackage{color}
\usepackage{rotating}
\usepackage[margin=0.9in,a4paper]{geometry}
\usepackage[table,xcdraw,dvipsnames]{xcolor}
\usepackage[utf8]{inputenc}
\usepackage{colortbl}
\usepackage[normalem]{ulem}
\usepackage{mathrsfs} 
\usepackage[version=4]{mhchem}
\usepackage{cancel}

\definecolor{nicered}{rgb}{0.7,0.1,0.1}
\definecolor{nicegreen}{rgb}{0.1,0.5,0.1}
\definecolor{red}{rgb}{1.0, 0, 0}

\newcommand{\bdm}{\begin{displaymath}}
\newcommand{\edm}{\end{displaymath}}
\newcommand{\bea}{\begin{eqnarray}}
\newcommand{\eea}{\end{eqnarray}}

\newcommand\aNLO{{\sc\small MadGraph5\_aMC@NLO}}

\def\be{\begin{equation}}
\def\ee{\end{equation}}

\definecolor{nicered}{rgb}{0.7,0.1,0.1}
\definecolor{nicegreen}{rgb}{0.1,0.5,0.1}
\definecolor{red}{rgb}{1.0, 0, 0}
\definecolor{niceblue}{rgb}{0,0,0.8}
\definecolor{red}{rgb}{1.0, 0, 0}
\hypersetup{colorlinks,citecolor= nicegreen,linkcolor= nicered,urlcolor=nicered}

\def\eq#1{{Eq.~(\ref{#1})}}

\def\fig#1{{Fig.~\ref{#1}}}

\def\sect#1{{Sect.~\ref{#1}}}
\def\sects#1#2{{Sects.~\ref{#1}--\ref{#2}}}

\def\gsim{\raise0.3ex\hbox{$\;>$\kern-0.75em\raise-1.1ex\hbox{$\sim\;$}}}
\def\lsim{\raise0.3ex\hbox{$\;<$\kern-0.75em\raise-1.1ex\hbox{$\sim\;$}}}

\def\mb[#1]{\mathbf{#1}}

\renewcommand{\bar}{\overline}

\definecolor{LightCyan}{rgb}{0.88,1,1}
\definecolor{piggypink}{rgb}{0.99, 0.87, 0.9}
\definecolor{applegreen}{rgb}{0.55, 0.71, 0.0}
\definecolor{darkpastelgreen}{rgb}{0.01, 0.75, 0.24}
\definecolor{green-yellow}{rgb}{0.68, 1.0, 0.18}

\newcommand{\beq}{\begin{equation}}
\newcommand{\eeq}{\end{equation}}
\newcommand{\beqa}{\begin{eqnarray}}
\newcommand{\eeqa}{\end{eqnarray}}

\begin{document}



\title{Hunting for a 17 MeV particle coupled to electrons}

\author{Luca Di Luzio}
\email{luca.diluzio@pd.infn.it}
\affiliation{\small \it Istituto Nazionale di Fisica Nucleare, Sezione di Padova, Via F.~Marzolo 8, 35131 Padova, Italy}

\author{Paride Paradisi}
\email{paride.paradisi@pd.infn.it}
\affiliation{Dipartimento di Fisica e Astronomia ``Galileo Galilei'',
Università degli Studi di Padova, Via F.~Marzolo 8, 35131 Padova, Italy}
\affiliation{\small \it Istituto Nazionale di Fisica Nucleare, Sezione di Padova, Via F.~Marzolo 8, 35131 Padova, Italy}

\author{Nudžeim Selimović}
\email{nudzeim.selimovic@pd.infn.it}
\affiliation{\small \it Istituto Nazionale di Fisica Nucleare, Sezione di Padova, Via F.~Marzolo 8, 35131 Padova, Italy}

\begin{abstract}
\noindent
We discuss a set of precision observables that can probe the existence of a light particle $X$ coupled to electrons in the mass range of 1--100~MeV.
As a case study, we consider the recent excess of $e^+e^-$ final-state events at $\sqrt{s} = 16.9$~MeV reported by the PADME collaboration. Interestingly, this mass is tantalizingly close to the invariant mass at which anomalous $e^+e^-$ pair production has previously been observed in nuclear transitions from excited to ground states by the ATOMKI collaboration.
For the scenario in which the new particle has a vector coupling to electrons, we show that the PADME excess is already in tension with constraints from the anomalous magnetic moment of the electron. Further improvements in the measurement of the electron $g$-2, together with upcoming results from PIONEER 
(searching for $\pi^+\to e^+ \nu X$)
and Mu3e 
(searching for $\mu^+ \to e^+ \bar\nu_\mu\nu_e X$), 
are expected to definitively probe this scenario in the near future.
We also explore alternative possibilities where the new particle has scalar, pseudoscalar, or axial-vector couplings.

\end{abstract}

\maketitle



\section{Introduction}
\label{sec:intro}

The PADME collaboration has recently reported an excess of $e^+e^-$ events peaked at $\sqrt{s} = 16.90$~MeV, with a moderate significance of $1.8\,\sigma$ (global) and $2.5\,\sigma$ (local)~\cite{Bossi:2025ptv,LDMA2025,LNFGenSem2025}. 
This energy scale matches the invariant mass at which anomalous $e^+e^-$ pair production has previously been observed, 
with high statistical significance, 
in nuclear transitions from excited to ground states in 
$^8$Be \cite{Krasznahorkay:2015iga,Krasznahorkay:2018snd}, 
$^4$He \cite{Krasznahorkay:2019lgi,Krasznahorkay:2019lyl,Krasznahorkay:2021joi,Krasznahorkay:2023sax,Krasznahorky:2024adr}, and $^{12}$C \cite{Krasznahorkay:2022pxs}
by the ATOMKI collaboration, 
as well as in an experiment conducted at 
the VNU University of Science involving $^8$Be \cite{Anh:2024req}.
Such a coincidence of scales significantly reinforces the case for a common underlying origin, potentially involving a hypothetical new boson, commonly referred to as $X_{17}$ (see e.g.~\cite{Feng:2016jff,Feng:2016ysn,Feng:2020mbt,Zhang:2020ukq,Barducci:2022lqd,Denton:2023gat,Alves:2023ree,Abdallah:2024uby}).

The MEG II experiment at PSI~\cite{MEGII:2024urz} also searched for the $X_{17}$ particle using the same $^7$Li($p$,\,$e^+e^-$)\,$^8$Be reaction employed by ATOMKI. While no significant signal was observed, the results are compatible with ATOMKI within $1.5\,\sigma$, and can be consistently included in a global fit \cite{Barducci:2025hpg}. Adding also the recent PADME result \cite{Bossi:2025ptv,LDMA2025,LNFGenSem2025} one obtains the best-fit value 
$m_X = 16.88(5)$ MeV \cite{Arias-Aragon:2025wdt}.  

Currently, there is no clear indication of the spin-parity nature of the $X_{17}$ particle. Scalar or pseudoscalar couplings to nucleons fail to simultaneously account for all observed nuclear transitions, while vector and axial-vector scenarios face tensions with other low-energy constraints on the $X_{17}$ couplings to nucleons (see e.g.~\cite{Zhang:2020ukq,Barducci:2022lqd,Hostert:2023tkg}). In contrast, the coupling to electrons, relevant only through the branching ratio of $X_{17}$ into $e^+e^-$, plays only a minor role in explaining the nuclear decay anomalies, but it can be directly tested by PADME \cite{Nardi:2018cxi,Darme:2022zfw,Arias-Aragon:2024qji,Bertelli:2025mil}.

While other beam-dump experiments in the $X_{17}$ mass range, such as NA64~\cite{NA64:2019auh,NA64:2021aiq}, 
could directly probe the $X_{17}$ coupling to electrons, potentially closing the entire available parameter space \cite{Beacham:2019nyx}, 
it remains important to further explore complementary and independent tests that could confirm or disprove the PADME excess.

In this work, we examine a set of observables sensitive to a new light particle coupled to electrons in the 1--100 MeV mass range. While exploring in more generality various coupling scenarios (scalar, pseudoscalar, vector, and axial-vector) we focus in particular on a 17 MeV state with vector couplings to electrons, motivated by the recent PADME analysis. Specifically, we assess current constraints and future sensitivities from the electron anomalous magnetic moment, exotic pion decays at the 
SINDRUM experiment~\cite{SINDRUM:1986klz,SINDRUM:1989qan}, as well as upcoming results from PIONEER~\cite{PIONEER:2022yag} and Mu3e~\cite{Mu3e:2020gyw}, which aim to search for exotic decays of the pion and muon, respectively. These experiments are expected to decisively test the above scenario in the near future.

The paper is structured as follows. In \sects{sec:obs}{sec:muon}, we introduce the relevant observables, providing a brief overview of the key formulae and the expected experimental sensitivities. Our results are presented in \sect{sec:results}, along with a description of the underlying phenomenological analysis. We conclude in \sect{sec:concl}, summarizing the main findings and outlining the prospects for testing the $X_{17}$ hypothesis through its coupling to electrons.

\section{Phenomenological setup}
\label{sec:obs}

Let us define the interaction Lagrangian of a new 
light state, $X$, with electrons as 
\beq
\label{eq:LeX}
\mathcal{L}_X \supset g_{eX}
\left(\bar e \,\Gamma_X e\right) X \, ,
\eeq
where $X=S,P,V,A$, and
$\Gamma_S = 1 $ if $S$ is a scalar, 
$\Gamma_P = i\gamma_5$ if $P$ is a pseudoscalar,
$\Gamma_V = \gamma_\mu$ if $V_\mu$ is a vector, 
and $\Gamma_A = \gamma_\mu\gamma_5$ if $A_\mu$ is an axial-vector.

In the following, we consider several observables that can probe the interaction of the $X$ particle with electrons, 
as described in \eq{eq:LeX}. 

For a vector coupling, the 
best-fit value associated with the $e^+e^-$ excess observed by the PADME collaboration is given by \cite{Bossi:2025ptv,LDMA2025,LNFGenSem2025}:
\begin{equation}
g_{eV} = 5.6 \times 10^{-4} \,, \quad 
m_V = 16.9 \ \text{MeV} \,.
\label{eq:best_fit}
\end{equation}
We will use this benchmark point to assess its compatibility with existing experimental constraints.

\section{Electron $\boldsymbol{g-2}$}
\label{sec:gme}

The anomalous magnetic moment of the electron, $a_e \equiv (g_e-2)/2$, has been commonly used to extract the value of the fine-structure constant, $\alpha$. However, recent improvements in atomic-physics experiments using Cesium (Cs) and Rubidium (Rb) interferometry have led to the following results for $\alpha$:
\begin{align}
\alpha({\rm Cs}) &= 1/137.035999046(27) \quad \text{\cite{Parker:2018vye}} \, , \\
\alpha({\rm Rb}) &= 1/137.035999206(11) \quad \text{\cite{Morel:2020dww}} \,,
\end{align}
showing a disagreement of $5.5\, \sigma$. Using the above determinations of $\alpha$ to predict the Standard Model (SM) value $a^{\rm SM}_e$ and comparing it with the latest experimental measurement of $a^{\rm exp}_e = (115\,965\,218\,059~\pm~13) \times 10^{-14}$~\cite{Fan:2022eto}, yields
the following values of $\Delta a_e \equiv a_e^{\rm exp} - a_e^{\rm SM}$:
\begin{align}
\label{eq:daeCs} 
(\Delta a_e)_{\rm Cs} &= (-102.0 \pm 26.4)\times 10^{-14} 
\, , \\
\label{eq:daeRb}
(\Delta a_e)_{\rm Rb} &= \,\,\,\,\,\,\,\,(33.8 \pm 16.1)\times 10^{-14} 
\,.
\end{align}
It is interesting to consider the dominant sources of error in the determinations of $(\Delta a_e)_{\rm Cs}$ and $(\Delta a_e)_{\rm Rb}$, which stem from the atomic measurement of $\alpha$ and the experimental measurement of $a^{\rm exp}_e$~\cite{DiLuzio:2024sps}: 
\begin{table}[ht!] 		
\begin{tabular}{|c|c|c|}
\hline
$\Delta a_e$ error source & 
Value &
\, \, \, \% of $\Delta a_e$ error \, \, 
\\
\hline
$\alpha({\rm Cs})$, $\delta a_{e}^{\alpha({\rm Cs})}$ & $22 \times 10^{-14}$ &  $70\%$ \\
$\alpha({\rm Rb })$, $\delta a_{e}^{\alpha({\rm Rb})}$ & $9 \times 10^{-14}$ &  $28\%$ \\
Experiment, $\delta a_e^{\rm exp}$ & $13 \times 10^{-14}$ &  $24\%\,({\rm Cs})/59\%\,({\rm Rb})$ \\
\hline
\end{tabular}
\label{table:aeErrors}
\vspace{-0.3cm}
\end{table}
\noindent
\vspace{8pt}

An improvement on $\delta a_e^{\rm exp}$ by a factor of $\sim 5$ is expected in the next few years.\footnote{G.~Gabrielse, private communication.} On the same timescale, new measurements of $\alpha({\rm Rb})$ and $\alpha({\rm Cs})$ will significantly reduce the systematic effects that were dominant in their previous measurements, hopefully resolving the current discrepancy.\footnote{S.~Guellati-Khélifa and H.~Mueller, private communications.}

Therefore, $a_e$ can be used as a precision test of the SM and its  extensions~\cite{Giudice:2012ms,Crivellin:2018qmi,Erdelyi:2025axy}.
The contribution of the boson $X$ to the electron $g$-2 
stemming from the Lagrangian of Eq.~(\ref{eq:LeX}) reads~\cite{Jegerlehner:2017gek,Athron:2021iuf}:
\begin{align}
\Delta a^X_e = \frac{g^2_{eX}}{4\pi^2} \frac{m^2_e}{m^2_X}\, L_X \, ,   
\end{align}
where, in the limit $m_X \gg m_e$, it turns out that
\begin{align}
L_S &= \ln\frac{m_S}{m_e} -\frac{7}{12} \,,\\
L_P &= -\ln\frac{m_P}{m_e} +\frac{11}{12} \,,\\
L_V &= \frac{1}{3} \,,~~
L_A = -\frac{5}{3} \,,
\end{align}
leading to the following numerical estimates:
\begin{align}
\!\!\!\Delta a^S_e &\!\approx 2.1 \!\times\! 10^{-11}\! \left(\frac{g_{eS}}{5.6 \!\times\! 10^{-4}}\right)^{\!2} \!\!
\left(\frac{17\,{\rm MeV}}{m_S}\right)^{\! 2}\!,  
\label{eq:gm2_predictionS}\\
\!\!\!\Delta a^P_e &\!\approx -1.9 \!\times\! 10^{-11}\! \left(\!\frac{g_{eP}}{5.6 \!\times\! 10^{-4}}\!\right)^{\!2} \!\!
\left(\!\frac{17\,{\rm MeV}}{m_P}\!\right)^{\! 2}\!,  \\
\!\!\!\Delta a^V_e &\!\approx 2.4 \!\times\! 10^{-12}\! \left(\frac{g_{eV}}{5.6 \!\times\! 10^{-4}}\right)^{\!2} \!\!
\left(\frac{17\,{\rm MeV}}{m_V}\right)^{\! 2}\!,  \\
\!\!\!\Delta a^A_e &\!\approx -1.2 \!\times\! 10^{-11}\! \left(\!\frac{g_{eA}}{5.6 \!\times\! 10^{-4}}\!\right)^{\!2} \!\!
\left(\!\frac{17\,{\rm MeV}}{m_A}\!\right)^{\! 2}\!\!,
\label{eq:gm2_predictionA}
\end{align}
where the PADME best-fit value of Eq.~(\ref{eq:best_fit}), which refers only to the vectorial case, has been assumed 
also for the other scenarios just as a reference. 

Comparing the results of Eqs.~(\ref{eq:gm2_predictionS}-\ref{eq:gm2_predictionA}) with
Eqs.~(\ref{eq:daeCs}) and (\ref{eq:daeRb}) we learn that the PADME result, see Eq.~(\ref{eq:best_fit}), is in a significant tension 
with the electron $g$-2 bound. 
Hereafter, for definiteness, we assume three benchmark scenarios: 1) $|\Delta a_e| \leq 10^{-12}$, where we inflated the current experimental errors in $\delta a^{\alpha(Cs)}_e$ and 
$\delta a^{\alpha(Rb)}_e$ to make Eqs.~(\ref{eq:daeCs}) and 
(\ref{eq:daeRb}) consistent, 2) $|\Delta a_e| \leq 10^{-13}$,
assuming a resolution of the current discrepancy in the measurements 
of $\alpha$ with a precision of $\mathcal{O}(10^{-13})$, and 3) 
$|\Delta a_e| \leq 10^{-14}$, which is the ultimate expected uncertainty on $\Delta a_e$ if both $\delta a^{\rm exp}_e$ and 
$\delta a^\alpha_e$ will improve by roughly one order of magnitude~\cite{DiLuzio:2024sps}.

\section{Pion decay}
\label{sec:pion}

Another way to detect the $X$ boson is through the charged pion decay $\pi^+\to e^+ \nu X$.\footnote{For a related analysis, including also vector and axial-vector couplings to quarks, see Ref.~\cite{Hostert:2023tkg}. The main focus here is 
on the electron coupling and the interplay of this observable with the PADME excess.} 
The SINDRUM experiment searched for $e^+e^-$ resonances in 
$\pi^+\to e^+ \nu X$, $X\to e^+ e^-$ with sensitivity to branching ratios of $\mathcal{O}(10^{-10})$~\cite{SINDRUM:1986klz,SINDRUM:1989qan}. In order to constrain the parameter space of our scenarios, we exploit the bounds by SINDRUM~\cite{SINDRUM:1989qan} given as a function of the $X$ mass.\footnote{The signal in the search was modeled using the differential decay rate for a light scalar, whose kinematic features closely resemble those of the longitudinal modes of vector bosons. As shown in~\cite{Hostert:2023tkg}, the small kinematic differences between the two cases are not expected to impact the interpretation of the resulting limits on spin-1 particles.}
If $X$ is invisible, a complementary limit can be derived from a search for $\pi^+\to e^+ \nu X$ at PIENU~\cite{PIENU:2021clt}.
However, in the parameter space of our interest, we find that $X$
decays promptly into the visible $e^+e^-$ channel and therefore 
the PIENU bounds do not apply in our case. 
In the future, charged pion decays into visible final states can 
be searched for at the proposed experiment PIONEER~\cite{PIONEER:2022yag}, planning to reach a sensitivity to
branching ratios of $\mathcal{O}(10^{-11})$. 

To assess the expected sensitivity on the coupling $g_{eX}$, it is useful to estimate the decay rate of $\pi^+\to e^+ \nu X$ in naive dimensional analysis (NDA), assuming $m_X \ll m_{\pi^+}$. 

When $X=S,P$, we find that $\Gamma(\pi^+\to e^+ \nu X)$ $\sim g^2_{eX} G^2_F m^3_\pi f^2_\pi |V_{ud}|^2/(4\pi)^3$, where $G_F$ is the Fermi constant, $f_\pi$ the pion decay constant, $V_{ud}$ the relevant CKM matrix element, and $1/(4\pi)^3$ stands for the three body phase space. Therefore, we find that
\begin{align}
\!\!\! 
\mathcal{B}(\pi^+ \!\to\! e^+ \nu X) \simeq 
\frac{\Gamma(\pi^+ \!\to\! e^+ \nu X)}{\Gamma(\pi^+ \!\to\! \mu^+ \nu)} 
\!\approx \frac{m^2_\pi}{m^2_\mu}\frac{g^2_{eX}}{16\pi^2}\,,
\label{eq:Br_SP}
\end{align}
where we used the standard pion decay rate $\Gamma(\pi^+\to \mu^+ \nu) \approx G^2_F m_\pi m^2_\mu f^2_\pi |V_{ud}|^2/4\pi$.
The current experimental bound $\mathcal{B}(\pi^+ \!\to\! e^+ \nu X) \lesssim 6 \times 10^{-10}$~\cite{SINDRUM:1986klz,SINDRUM:1989qan} by SINDRUM is therefore expected to translate into a bound on the coupling $g_{eX} \lesssim \, \text{few} \times 10^{-4}$.

The exact analytic expressions for the differential decay rates of leptonic meson decays accompanied by the emission of an axion-like particle have been derived in Refs.~\cite{Aditya:2012ay,Gallo:2021ame,Altmannshofer:2022ckw}. We have verified that~\aNLO~\cite{Alwall:2014hca} accurately reproduces these known results, and we have further utilized it to obtain the corresponding predictions for the scalar and (axial-)vector cases. For the representative benchmark of $m_X = 17$~MeV, and normalising to the SINDRUM sensitivity $\mathcal{B}(\pi^+\to e^+ \nu X) \lesssim 6\times10^{-10}$ at this $X$ mass, we find that
\begin{align}
\!\! \frac{\mathcal{B}(\pi^+\to e^+ \nu X)}{6\times10^{-10}} &\approx 3.9 \left(\frac{g_{eX}}{5.6 \!\times\! 10^{-4}}\right)^{\!2} \, , 
\end{align}
when $X=S,P$, in good agreement with NDA in Eq.~\eqref{eq:Br_SP}. 

On the other hand, if $X=V,A$ is a gauge boson coupled to a non-conserved current in an effective field theory where gauge invariance is restored at a scale $\Lambda$, its longitudinal mode can participate in processes at energies $E<\Lambda$, leading to enhancements by factors of $\mathcal{O}(E^2/m_X^2)$.

This can be readily illustrated by considering the amplitude for $X$ emission in $\pi^+\to e^+ \nu X$, arising from the effective Lagrangian
\begin{align}
    \mathcal{L}&\supset \sqrt{2}G_F f_\pi V_{ud}^*\, (\partial_\mu \pi^+)\,\bar\nu_e \gamma^\mu {\rm{P}_L} e \label{eq:Lag_pi+decay}\\
    & + g_{eX} X_\mu \left(Q_e\, \bar e \gamma^\mu {\rm{P}_L} e + Q_\nu\, \bar \nu_e \gamma^\mu {\rm{P}_L} \nu_e\right)\,,\nonumber
\end{align}
where $Q_e$ and $Q_\nu$ denote the $U(1)_X$ charges of the electron and electron neutrino, respectively. For the momentum configuration specified by $\pi^+(p_1)\to e^+(p_2) \nu_e(p_3) X(p_4)$, the amplitude reads
\begin{align}
    \mathcal{M} &= \sqrt{2}G_F f_\pi V_{ud}^*\, g_{eX} \bar u(p_3) \Gamma^\mu v(p_2) \epsilon_\mu^*(p_4)\,, 
\end{align}
with
\begin{align}
    \Gamma^\mu &= (Q_e-Q_\nu) \gamma^\mu {\rm{P}_L}\label{eq:amplitude}\\
    &+Q_e \frac{m_e^2}{p_{24}^2-m_e^2}\gamma^\mu {\rm{P}_L} + Q_\nu \gamma^\mu \frac{\slashed{p}_{34}}{p_{34}^2}m_e {\rm{P}_R}\,,\nonumber
\end{align}
where $p_{ij} = p_i+p_j$. 

In the case of weak-violation—where the components of an $SU(2)_L$ lepton doublet carry different charges under $U(1)_X$, the first term in Eq.~\eqref{eq:amplitude} dominates the amplitude, leading to a large enhancement of $\mathcal{O}(m_\pi^4/m_X^2 m_e^2)$ in $\mathcal{B}(\pi^+\to e^+\nu X)$, and thus to strong constraints on light spin-1 mediators~\cite{Dror:2017nsg,Dror:2020fbh,Hostert:2023tkg}.

Importantly, this enhancement is absent in the weak-conserving case $Q_e=Q_\nu$, or more generally when the current is conserved, as in the case of a dark photon or $U(1)_{X=B-L}$ leading to significantly weaker bounds. Although such scenarios are not captured by the simplified model in Eq.~\eqref{eq:LeX}, their limits can be understood in terms of the $Q_e=Q_\nu$ case in Eq.~\eqref{eq:Lag_pi+decay}. Moreover, introducing couplings of $X$ to additional SM fermions would necessitate the inclusion of further observables.

Therefore, when $X=V,A$ and $Q_\nu=0$, we find
\begin{align}
\!\! \frac{\mathcal{B}(\pi^+\to e^+ \nu X)}{6\times10^{-10}} &\approx 87 \left(\frac{Q_e\,g_{eX}}{5.6 \!\times\! 10^{-4}}\right)^{\!2} \, , 
\end{align}
for $m_X=17$ MeV, again in good agreement with NDA after taking into account the enhancement factor $\mathcal{O}(m_\pi^2/m_X^2)$ discussed above. Finally, when $Q_e=Q_\nu$, we have that
\begin{align}
\!\! \frac{\mathcal{B}(\pi^+\to e^+ \nu X)}{6\times10^{-10}} &\approx 6.5\times10^{-3}\left(\frac{Q_e\,g_{eX}}{5.6 \!\times\! 10^{-4}}\right)^{\!2} , 
\end{align}
demonstrating the chiral suppression evident in the amplitude in Eq.~\eqref{eq:amplitude}.

\section{Muon decay}
\label{sec:muon}

Muon decays could also be exploited to probe a light particle $X$ 
coupled to electrons \cite{Echenard:2014lma,Knapen:2023iwg,Knapen:2024fvh}. In particular, the upcoming Mu3e experiment at PSI searching for lepton flavour violation (LFV) in $\mu^+\to e^+ e^- e^+$ will benefit from an intense muon beam, producing and stopping $\mathcal{O}(10^{15})$ muons during Phase I, with an order of magnitude improvement in Phase II~\cite{Mu3e:2020gyw}. 

The signature of interest in our analysis differs from the LFV process $\mu^+ \to e^+ e^- e^+$. Instead, we focus on the scenario in which a light particle $X$ is radiated from a final-state positron in the decay 
$\mu^+ \to e^+ \bar\nu_\mu\nu_e X$, with $X$ subsequently decaying into an $e^+e^-$ pair. As a result, the invariant mass of the electron and one of the positrons in the final state satisfies the relation $(p_{e^+}+p_{e^-})^2 = m^2_X$. The experimental signature of this process is therefore characterized by a narrow bump in the invariant mass spectrum of $e^+ e^-$ pairs, appearing on top of the SM background from the five-body decay $\mu^+ \to e^+ e^- e^+ \bar{\nu}_\mu \nu_e$. 

In addition to analyzing the invariant mass distributions of the $e^+ e^-$ pairs for resonance searches, Ref.~\cite{Knapen:2023iwg} investigated the impact of incorporating angular observables. This inclusion was shown to enhance the expected experimental sensitivity to the (pseudo)scalar-electron coupling $g_{eX}$, with $X=S,P$, by a factor of $\mathcal{O}(1)$. For the case of such light states, we adopt the projected sensitivities from Ref.~\cite{Knapen:2023iwg} and present them in Fig.~\ref{fig:resultsALL} (upper panel), using different shades of blue to indicate the corresponding Mu3e phase.
In contrast, for the case of (pseudo)vector mediators, Ref.~\cite{Knapen:2023iwg} does not provide projections assuming couplings exclusively to electrons. To address this, we adopt the projected sensitivity to ${\rm Br}(\mu^+ \to e^+ \nu_e \bar\nu_\mu V, V\to e^+ e^-)$ from Ref.~\cite{Perrevoort:2018okj} and compute the corresponding branching ratio using \aNLO, considering only radiation from the positron. The resulting constraints on the coupling $g_{eX}$, with $X=V,A$ are shown in Fig.~\ref{fig:resultsALL} (lower panel).

In comparison to pion decay experiments such as SINDRUM and PIONEER, the projected sensitivity of Mu3e is generally weaker if $X$ couples only to electrons. This difference arises from the chirality suppression present in the SM background $\pi^+\to e^+\nu_e \gamma^*, \gamma^*\to e^+ e^-$, which does not apply to muon decays. However, if $X=V,A$ couples in a weak-conserving manner with $Q_e=Q_\nu$ in Eq.~\eqref{eq:Lag_pi+decay}, or more generally to a conserved current, the same chiral suppression also applies to the signal in pion decays, and Mu3e exhibits significantly better sensitivity in comparison.

Additionally, we carried out an independent sensitivity study using the {\sc\small FeynRules} –{\sc\small Ufo}–{\aNLO} toolchain~\cite{Alloul:2013bka,Degrande:2011ua,Alwall:2011uj,Alwall:2014hca} to simulate both signal and background processes for $\mu^+ \to e^+ e^- e^+ \nu_e \bar\nu_\mu$. Adopting a fixed mass resolution of $0.5$ MeV across the relevant invariant mass range as in~\cite{Knapen:2023iwg}, we perform a bin-by-bin analysis over the mediator mass range  $m_X \in [2\,,100]$~MeV. For each bin, we apply Poisson statistics to derive upper bounds on the coupling $g_{eX}$. Given the large number of background events $B$ per bin, the Gaussian approximation is applicable, and we quantify sensitivity using $S/\sqrt{B}\simeq 1.64$, which corresponds approximately to a 95\% confidence level (CL), under the assumption that statistical uncertainties dominate. We confirm the results of~\cite{Perrevoort:2018okj,Knapen:2016moh} up to $\mathcal{O}(1)$ differences attributable to detector effects.

\section{Results}
\label{sec:results}

\begin{figure*}[t!]
\centering
\includegraphics[width=0.49\textwidth]{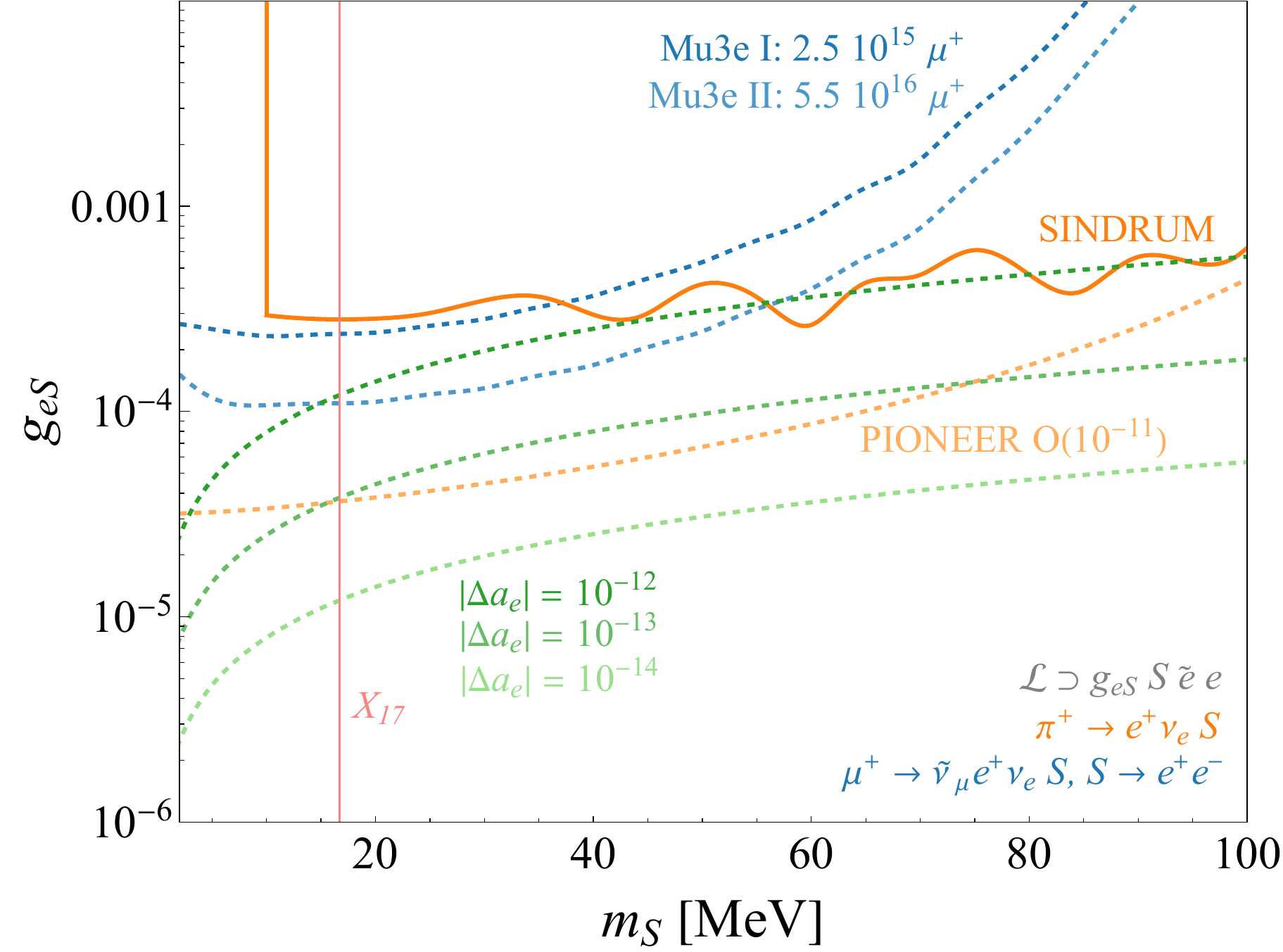} \ \
\includegraphics[width=0.49\textwidth]{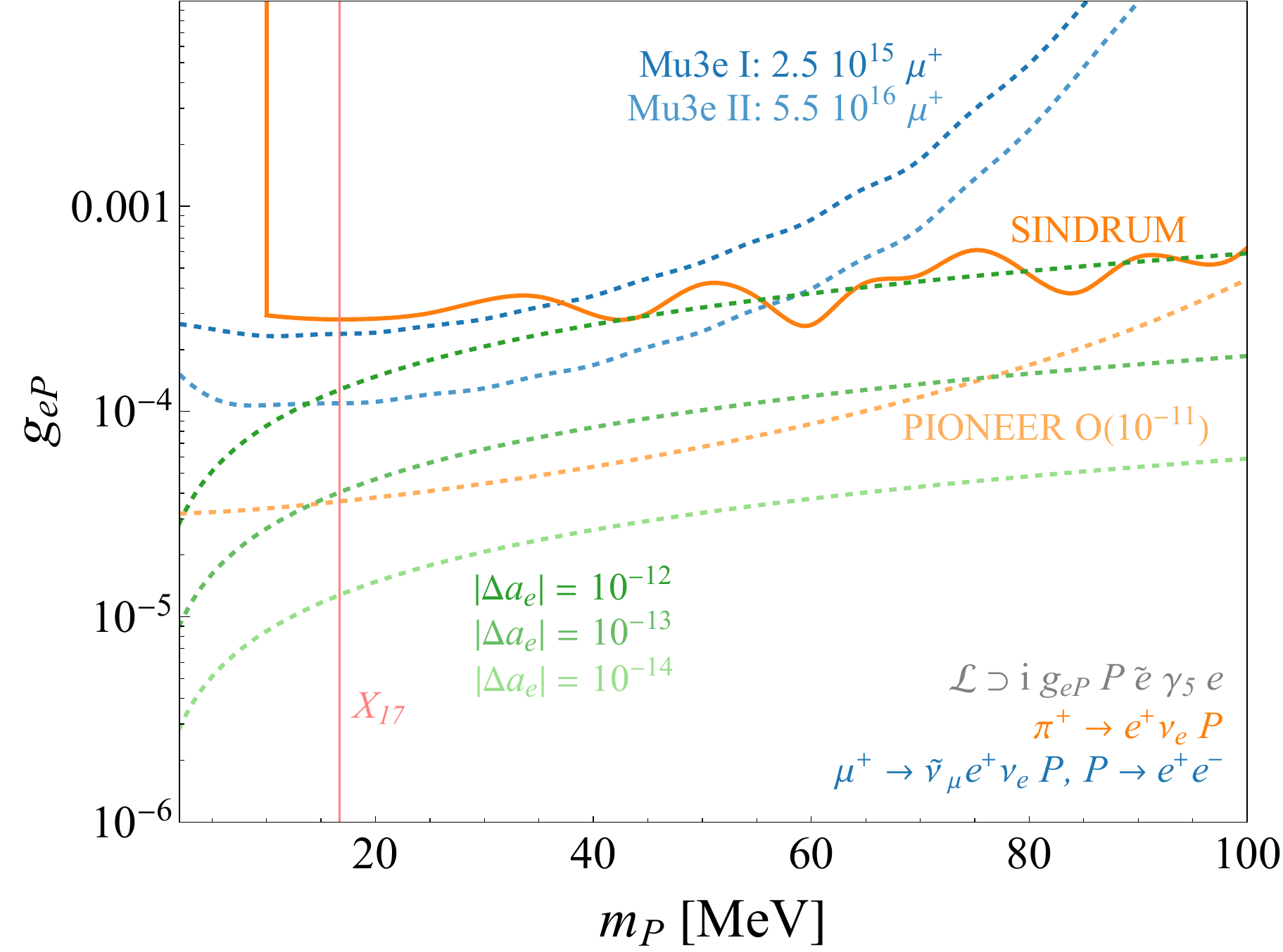} \\
\vspace{0.5cm}
\includegraphics[width=0.49\textwidth]{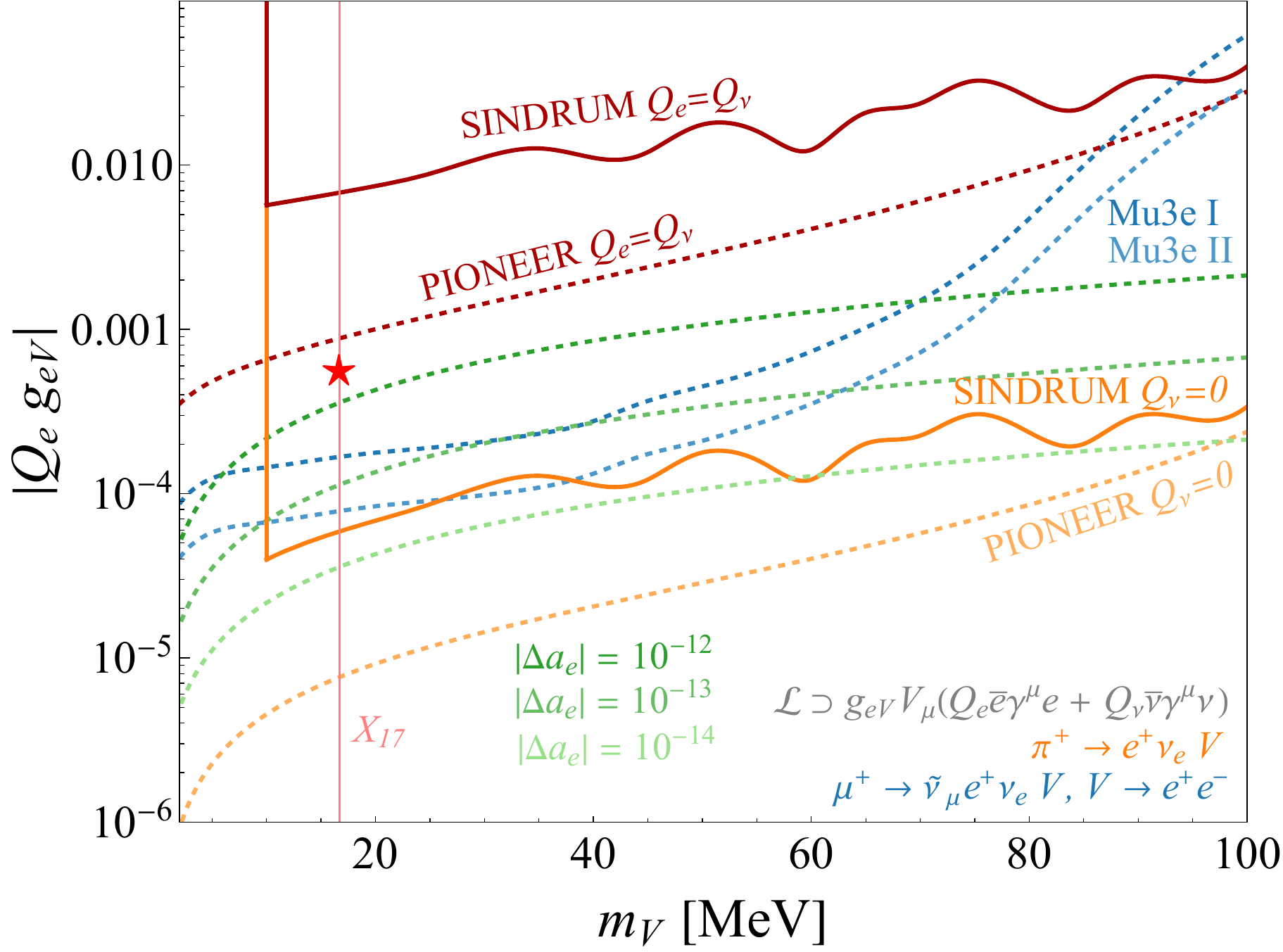} \ \
\includegraphics[width=0.49\textwidth]{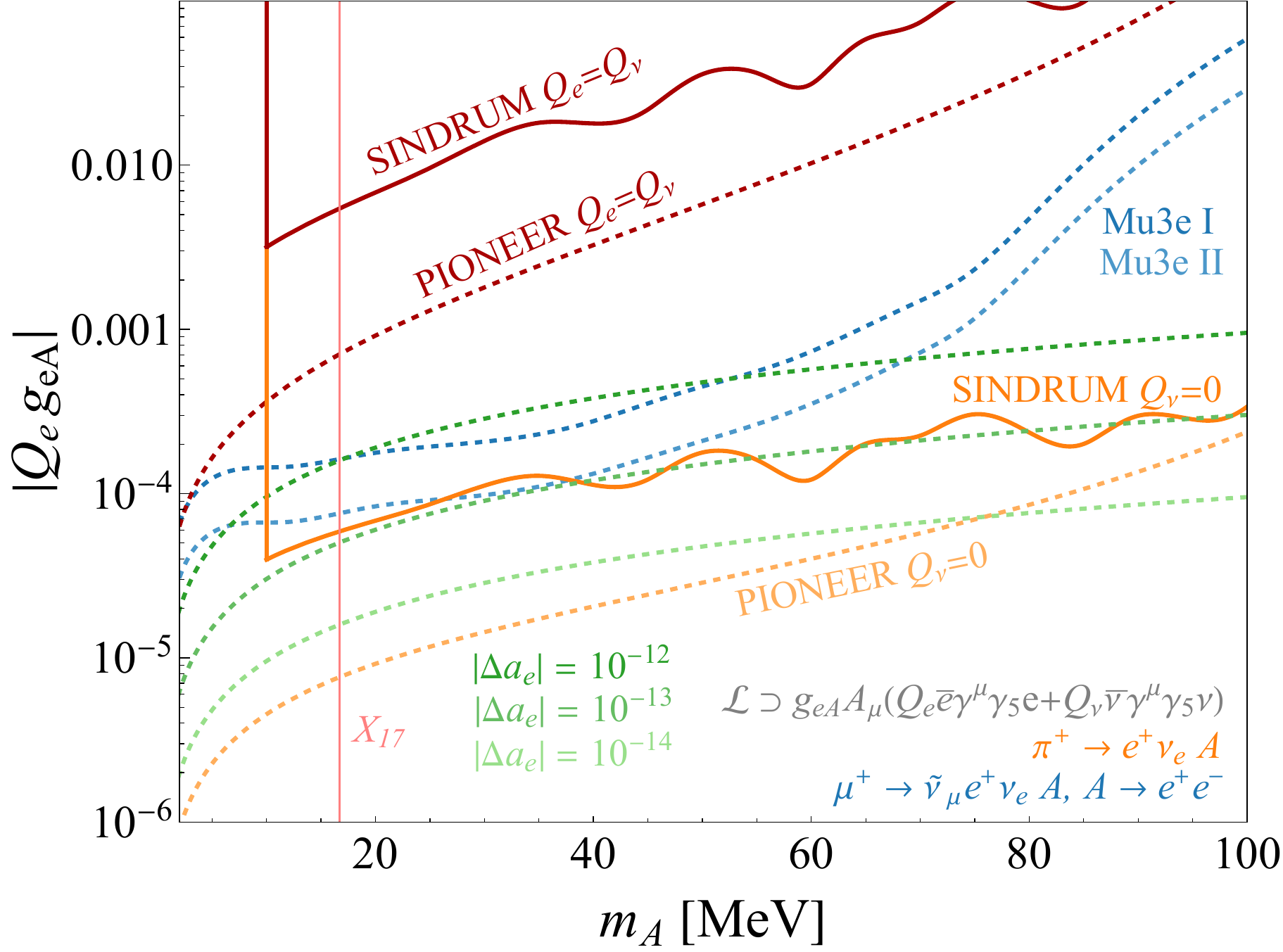}
\caption{Present constraints (solid lines) and expected sensitivities (dashed lines) for a light $X=S,P,V,A$ particle coupled to electrons in the 1--100 MeV 
mass range. The vertical, red line at $m_X = 16.9$ MeV denotes 
the $X_{17}$ benchmark, with the red star (in the vector case) 
representing the PADME best-fit value $Q_e\,g_{eV} = 5.6 \times 10^{-4}$. The burgundy (orange) lines illustrate the constraints from the exotic pion decays under the hypothesis that $X=V,A$ couples to a conserved (non-conserved) current.}
\label{fig:resultsALL}
\end{figure*}

The results of our analysis are displayed in \fig{fig:resultsALL}, which shows the present constraints (solid lines) and expected sensitivities (dashed lines) for a light $X=S,P,V,A$ particle coupled to electrons in the 1--100 MeV mass range.

When the light particle is a spin-0 state, we find the limits to be comparable between the scalar ($X=S$ in the upper-left panel of Fig.~\ref{fig:resultsALL}) and the pseudoscalar ($X=P$ in the upper-right panel). In both cases, the most stringent constraints arise from measurements of the electron $g$-2 (dashed lines in different shades of green), with the precision in $|\Delta a_e|$ of $\mathcal{O}(10^{-14})$ allowing to probe the (pseudo)scalar couplings at the level of $g_{eX}\lesssim 10^{-5}$.  A more moderate precision of $|\Delta a_e| \simeq \mathcal{O}(10^{-13})$ results in bounds which overlap to a good extent with the projections at the proposed PIONEER experiment which aims to reach sensitivities in radiative pion decays at the level of $\mathcal{B}(\pi^+ \to e^+ \nu X)\lesssim \mathcal{O}(10^{-11})$ (dashed orange line). Interestingly, the current bounds from the SINDRUM experiment (solid orange line) searching for the neutral particle emitted in $\pi^+$ decay approximately correspond to the sensitivity in $|\Delta a_e| \simeq \mathcal{O}(10^{-12})$ in the mass region above 40 MeV. At present, these constitute the most sensitive probes of new physics scenarios coupled to electrons in the 1–100 MeV mass range. Future muon experiments are expected to reach a sensitivity of $g_{eX}\lesssim 10^{-4}$ through the peak search in the invariant mass spectrum of $e^+e^-$ pair originating from a narrow resonance $X$, supplemented by the angular observables constructed from the $e^\pm$ momenta. The projected limits from the Mu3e experiment are shown as dashed blue lines, accounting for two planned phases:  Phase I, with $\mathcal{O}(2.5\times10^{15})$ muon decays, and Phase II, with an expected total of $\mathcal{O}(5.5\times10^{16})$ events. 

When the light particle is a spin-1 state, the limits are overall more stringent than in the spin-0 case. This is particularly evident for the constraints from exotic $\pi^+$ decays, as set by SINDRUM (solid orange) and projected PIONEER (dashed orange) experiments when $Q_e\ne Q_\nu$ in Eq.~\eqref{eq:Lag_pi+decay}, and we show the case with $Q_\nu=0$. These stronger bounds arise due to the absence of the chiral suppression in the weak-violating or non-conserved current case (see Sec.~\ref{sec:pion}), and the additional enhancement from the emission of the longitudinal mode, which contributes a factor of $\mathcal{O}(m_\pi^2/m_X^2)$ to $\mathcal{B}(\pi^+\to e^+ \nu X)$.  This enhancement is especially significant in the low-mass region when compared with the spin-0 scenarios. Indeed, these pion decay limits represent the most stringent constraints, followed by those from projections on the precision of $\Delta a_e$ (dashed green lines), reversing the pattern observed in the spin-0 case. However, when $Q_e=Q_\nu$, the aforementioned enhancements are absent, and the decay remains chirally suppressed, resulting in significantly weaker constraints from exotic pion decays probed by the SINDRUM and PIONEER experiments, shown by the solid and dashed burgundy lines in Fig.~\ref{fig:resultsALL}, respectively.

Regarding the experiments utilizing intense muon beams: in Phase II of the Mu3e experiment, the spin-1 scenario is expected to be probed down to $g_{eX} \simeq 8\times 10^{-5}$ (for $Q_e=1$ and $Q_\nu=0$)\footnote{For the $Q_e=Q_\nu$ case, the difference in the $g_{eX}$ limit is $\mathcal{O}(1)$ and is therefore omitted from Fig.~\ref{fig:resultsALL} to avoid further clutter.} at the benchmark mass of $m_X=17$ MeV, compared to $g_{eX} \simeq 10^{-4}$ in the spin-0 case. 
Furthermore, the axial-vector scenario ($X=A$, shown in the lower-right panel of Fig.~\ref{fig:resultsALL}) is more tightly constrained by the electron $g$-2 than the vector case ($X=V$, shown in the lower-left panel).

Remarkably, the PADME best-fit value of $Q_e\, g_{eV} =5.6 \times 10^{-4}$ (indicated by the red star in Fig.~\ref{fig:resultsALL}) is already 
in tension with $\Delta a_e$ and excluded by SINDRUM when $Q_e \ne Q_\nu$. In contrast, when the new physics is weak-conserving with $Q_e=Q_\nu$, or $X_\mu$ couples to a conserved current, the PADME best-fit coupling is still allowed, but clearly within reach of upcoming experiments shown in Fig.~\ref{fig:resultsALL}.

\phantom{Although we focused on the vector case motivated by the recent PADME analysis, we have also examined scalar, pseudoscalar, and axial-vector couplings, which may be relevant 
for other $X_{17}$ scenarios or in other contexts. 
As further data from PADME and related experiments becomes available, the framework explored here will remain a useful tool in scrutinizing the $X_{17}$ interpretation. Although we focused on the vector case motivated by the recent PADME analysis, we have also examined scalar, pseudoscalar, and axial-vector couplings, which may be relevant 
for other $X_{17}$ scenarios or in other contexts. 
As further data from PADME and related experiments becomes available, the framework explored here will remain a useful tool in scrutinizing the $X_{17}$ interpretation. }

\section{Conclusions}
\label{sec:concl}

\vspace{-0.4cm} In this work, we have tested the hypothesis of a new 17 MeV boson coupled to electrons, commonly referred to as $X_{17}$, by employing a set of observables complementary to those probed by PADME and other beam-dump experiments. 
We have shown that the $e^+e^-$ PADME excess 
is already in tension with the electron $g$-2, predicting a correlated signal with $|\Delta a_e| > 10^{-12}$. The same holds for the SINDRUM experiment that effectively excludes the PADME excess, except in scenarios where $X_{17}$ is a spin-1 boson coupled to electrons and electron neutrinos with equal strength, or more generally to a conserved current, as in the case of a dark photon or $U(1)_{B-L}$.

We have analyzed a set of observables sensitive to the coupling of a light particle to electrons in the 1--100 MeV mass range, identifying regions of parameter space that are currently allowed and those soon to be probed. In particular, future measurements from Mu3e, PIONEER, and improved determinations of the electron $g$-2 will be able to definitively test the $X_{17}$ scenario in the near future.

Although we focused on the vector case motivated by the recent PADME analysis, we have also examined scalar, pseudoscalar, and axial-vector couplings, which may be relevant 
for other $X_{17}$ scenarios or in other contexts. 
As further data from PADME and related experiments becomes available, the framework explored here will remain a useful tool in scrutinizing the $X_{17}$ interpretation. 

\phantom{Although we focused on the vector case motivated by the recent PADME analysis, we have also examined scalar, pseudoscalar, and axial-vector couplings, which may be relevant 
for other $X_{17}$ scenarios or in other contexts. 
As further data from PADME and related experiments becomes available, the framework explored here will remain a useful tool in scrutinizing the $X_{17}$ interpretation. Although we focused on the vector case motivated by the recent PADME analysis, we have also examined scalar, pseudoscalar, and axial-vector couplings, which may be relevant 
for other $X_{17}$ scenarios or in other contexts. 
As further data from PADME and related experiments becomes available, the framework explored here will remain a useful tool in scrutinizing the $X_{17}$ interpretation. }

\section*{Acknowledgments}
We thank Simon Knapen and Andre Schöning 
for useful communications.
This work received funding by the INFN Iniziative Specifiche AMPLITUDES and APINE and from the European Union’s Horizon 2020 research and innovation programme under the Marie Sklodowska-Curie grant agreements n.~860881 – HIDDeN, n.~101086085 – ASYMMETRY. 
This work was also partially supported by the Italian MUR Departments of Excellence grant 2023-2027 “Quantum Frontiers”.
The work of LDL and PP is supported by the European Union – Next Generation EU and by the Italian Ministry of University and Research (MUR) via the PRIN 2022 project n.~2022K4B58X – AxionOrigins. 


\bibliographystyle{apsrev4-1.bst}
\bibliography{bibliography}

\end{document}